# Aging in gaseous photodetectors

T. Francke, V. Peskov

Royal Institute of Technology, Stockholm, Sweden

**Abstract**

This paper describes the present status of the aging studies in various photosensitive detectors. New experimental data are presented on aging of trimethilamine (TMA) and ethilferrocene (EF) photosensitive vapors as well as on CsI and SbCs photocathodes. A new explanation of the CsI photocathodes aging process based on solid –state physics theory is given.

Finally, based on our studies, a general conclusion was made that thin polymer depositions on the detector's cathode due to the aging can provoke breakdowns through the Malter-type mechanism (or more precisely- an explosive field emission mechanism).

## I. Introduction

An interest to the photosensitive gaseous detectors appeared after publications [1,2] in which first photosensitive wire chamber were described. These detectors were flushed with benzene or tolyene vapors having a relatively low ionization potential. UV photons penetrating through a transparent window caused photoionization and this created primary photoelectrons, which then triggered avalanches. Nowadays, such types of detectors are widely used in many experiments. The reasons of their success are very simple: they have quantum efficiency (QE) in the UV region comparable or even higher than the best vacuum photomultipliers. At the same time they are position sensitive, cheap and simple. Of course they are not free from some drawbacks. One of the drawbacks is aging; another is the fact that their spectral sensitivity is limited by the ionization potential. The vapors with the lowest ionization potential $E_i$ used so far were: TMA ($E_i$=7.25 eV) [3], EF ($E_i$=6.08 eV) [4], TEA ($E_i$=7.15 eV) [5], TMAE ($E_i$=5.28 eV) [6]. Unfortunately in the last ten years or so there has been no progress in finding new vapors with $E_i$ lower than EF or TMAE and compatible for gaseous detectors.

This is why an interest arose to an alternative approach –to use solid photocathodes in gaseous detectors (see [7] and references therein). The threshold of the spectral sensitivity of such detectors is determined by the material work function, which can be very low. However such detectors could be competitive with those filled with photosensitive vapors only if they are stable in time and if their aging properties are satisfactory.

One should note that there are much more applications of photosensitive materials (vapors and solid photocathodes) that just in detecting UV photons. For example, EF vapors were used long time ago for plasma studies in so-called x-ray cluster counters in which individual primary electrons produced by x-rays were counted [4,8,9]. This approach was further developed by Breskin's group [10]. TMA, TEA and TMAE are widely used [11] for optical readout of gaseous detectors so called light chambers (optical readout [12]). Most of the solid photocathodes are also good secondary electron emitters and can be used in the detection of charged particles and x-ray photons [13]. The successes of all these applications depend very much on good aging properties of the materials used.

In this paper we will review the present status of aging studies in photosensitive detectors. New, yet unpublished results on this subject will be presented as well as a new interpretation of the CsI and other solid photocathodes aging.

Finally, a general conclusion will be stated that thin polymer depositions on the detector's cathode due to the aging can provoke breakdowns through the Malter-type mechanism.

**II. What is known about photosensitive materials aging?**

**II-1. Photosensitive vapors**

To update our knowledge the systematic aging studies were done only for wire chambers filled with TMAE and TEA vapors (see Fig1). It was found that the aging rate depends very much on system cleanness, anode wire diameter and only slightly on the composition of the carrier gas.

The analysis of polymer products on the anodes and the cathodes of the wire chambers were done in the past only in the case of TMAE [14]. It was found that on the anode wires the main depositions are not oxidized TMAE, but rather a tertiary amide. It remains unclear however, what fragments of TMAE react with oxygen to form tertiary amide. There have also been findings of some unidentified heavy organic compounds with atomic masses of 202, 110 and 100. On the cathode wires the following depositions were found: carbon in a form of graphite and small fractions of amides (N-C=O), silicone

[R$_2$SiO]$_n$ and amines NR$_3$ (R=C,H) (the silicon probably comes from the oil bubbler). The elements ratio was as follows:

83%C, 8.2%O, 5,9%N, 2.7%Si.

Thus many types of organic species coated the cathode. From the first glance it looks as if at least part of the problem comes from contaminations (for example Si).

The other important consequence of the aging was the Malter-type effect- a sporadic burst of electron emission from the cathode. The role of this effect in the detector's operation will be discussed at the end of this paper.

**II.2 Solid photocathodes**

Various photocathodes were used for various applications. For example, a wire chamber with a CuI photocathode was used successfully for plasma diagnostics [4]. However, the most popular today is CsI photocathode [15]. There are two main reasons for this: it has high QE (see Fig.2) and allows a short exposure to air, which is very convenient in the process of assembling the detector. Unfortunately, the aging data published so far are very contradictory [16]. This great spread of data is probably because of the fact that aging depends on cleanness, exposure to air and water vapors and other poorly controlled parameters.

One should note that the explanation of the CsI aging mechanism is not at all satisfactory. Indeed, the present theory is based on an assumption that CsI is dissociated under the UV radiation or ion bombardment (see for example [16]):

hv+CsI → Cs$^+$+I+e

A$^+$ +CsI → A+CsI$^+$---Cs$^+$+I

This model is certainly correct for CsI molecules. However in the case of the CsI in crystalline form, the aging mechanisms are different. To top it all, in most gaseous detectors operating at 1atm the positive ions do not have enough energy to dissociate the CsI.

**III. Experimental set up for aging tests in ultra clean conditions**

The main conclusion, which can be derived from previous studies, is that aging depends on the experimental set up cleanness. In this work we performed aging studies of several photosensitive materials, which were used in some our applications: TMA, EF, CsI, SbCs in the cleanest possible conditions.

One should note that it is very difficult (and expensive) to ensure an extremely clean gas system when it is continuously flushed with the gas. The main problem comes from the continuos outgassing of the materials used for the design of the gas system, test chamber and the detector itself. It is very difficult and takes a long time to outgas the experimental set up operating in a gas-flushing mode. It is much more efficient to outgas the set up, when it is pumped and heated to a reasonably high temperature and then use the detector in a sealed mode. This is why in this work we chose the later approach. Our experience show that if one tries to use the same system (prepared as described above) in a gas flushed mode, then it would be more difficult to ensure the same level of cleanness as it would be possible with a sealed detector. For example, an additional contamination comes from the oil bubbler, which brings Si-based contamination to the system

**III-1. Set up for EF and TMA aging studies**

Our experimental set up for the photosensitive vapor aging studied is presented in fig.3a. It is basically a station for the production of sealed single-wire counters filled with photosensitive vapors. It was extensively used in the past for manufacturing sealed detectors for plasma diagnostics. The station includes a single wire counter (SWC) itself, a heating and pumping system and a gas system. The cylindrical cathode of the SWC (diameter of 15 mm) was made of stainless steel, the anode wire of two different diameters (0.05 mm and 0.2 mm) made of molybdenum; the anode-cathode dielectric interface made of the glass. The detector had a LiF entrance window diameter of 2 mm. Before filling the gas and sealing itself, the SWC was pumped to a vacuum of ~$10^{-6}$ Torr and heated to 150°C for several days. The gas system contained two small glass bottles: an empty one and another one filled with a photosensitive liquid (EF or TMA) and a glass cylinder with a spectroscopic clean Ar, Kr or Xe (99.999% cleanness level). The EF and TMA were chemically cleaned at the Chemical Lab of the Moscow State University (D. Lemenovsky lab). In additional, before introducing in to the SWC, the liquids were distilled 6-10 times by "cryo" –transfer liquids from one glass bottle to another and pumped.

**III-2. Set up for CsI and SbCs photocathodes aging studies**

In the past, for the testing of solid photocathodes we used a system that allowed performing of the photocathodes aging tests without exposure to air [7]. This was achieved by using manipulators, which allowed the photocathodes to be transferred from an evaporation system to the test chamber. The test chamber was then flushed by a gas at

1atm. This allowed us to achieve rather good aging characteristics [7]. However, our experience with solid photocathodes shows that the best stability and aging characteristics can be achieved only in a compact sealed chamber. This is because in a system flushed by a gas or in a continuously pumped system, there is continuous delivery and accumulations of impurities on the photocathode surface. This is why in the present work we performed aging tests in sealed detectors only [17]. The experimental set up for these measurements is presented in Fig.3b. It contains a sealed detector with a solid photocathode, an UV source (a Hg lamp) and a reference detector: a photomultiplier or a photodiode monitoring the intensity of the UV radiation. To avoid any aging effect in the reference detector the intensity of the UV flux reaching the detector was attenuated by several orders of magnitude.

To separately check contributions to the aging process the light itself and the action of the light together with positive ions bombardments, we performed measurements in vacuum and the gas atmosphere with a small multiplication (factor of 5-10). In additional some control measurements were done with vacuum and gas filled Hamamatsu photodiodes: R1187, R1187 (filled with Ar at pressure 1 atm by a technique described in [18]), R414, R250.

**III-3 Experimental set-up for the aging studies of CsI secondary electron emitters**

As it was demonstrated in our earlier work [19] a thin gap parallel-plate chamber with a secondary electron sensitive layer allows to achieve very good position resolution for x-rays better than 50 μm in a simple counting mode [19]. This approach may find large-scale application in a digital imaging technique [20]. However, aging properties of such emitter should be demonstrated. In this work we performed some preliminary aging studies of the CsI converter under the simultaneous action of x-rays and bombardment of positive ions from the gas avalanches. Our set up is shown schematically in Fig.3c. It contains a test parallel –plate avalanche chamber (PPAC) with a stainless steel cathode coated by the CsI layer, a reference PPAC with the stainless steel cathode but without any CsI coating and an X-ray gun (6-30 keV). Both PPACs were flushed by Xe (40%)+Kr(49%)+$CO_2$(20%) gas mixture at 1atm. The x-ray flux hit the cathodes of both PPACs at shallow angles of about $10^0$. In contrast to our previous studies, in this work we used a uniform (not a porous) CsI layer of 0.4 μm thickness. The coating was done at CERN (Braem Lab). During the transfer to the test chamber the cathode was exposed to air for a few min, but during this exposure the area around the photocathode was continuously flushed by Ar. After the installation to the test camber it was pumped and heated for 24 hours. Only after that a working mixture was introduced.

**IV. Results**

**IV-1. Vapors**

The results of aging tests obtained with TMA and EF are presented in Fig.1. For comparison, on the same figure are presented some published data for TEA and TMAE vapors [9]. One can see that the aging properties of EF and TMA are more superior. Since TMA and TEA are chemically similar, one can speculate that the difference in aging properties is due to the cleanness.

**IV-2. CsI and SbCs photocathodes**

At low accumulated total charges of ~mC/cm$^2$ the results obtained with our sealed detectors and Hamamatsu photodiodes were similar: no aging effect was observed at all. Tests with Hamamatsu photodiodes were performed to much greater accumulated total charge: up to 30mC/cm$^2$. The results are presented in Fig. 4 (CsI) and Fig. 5 (SbCs), showing the measured photocurrent as function of irradiation time. Note that some degradation of the QE observed at 10 th day of continuous light illumination was reversible: after blocking the light for a day the QE returned to its original value. Note also that heating of these photocathodes does not exhibit any QE enhancements. This is in contrast to the photocathodes exposed to air [21].

To our knowledge the published results on photocathode aging in sealed detectors are very poor (see for example [10,22]) and do not allow any quantitative comparison to be preformed with our data.

**IV.3 Aging of the CsI x-rays converter**

Results of the CsI converter aging test under a combined action of x-rays and avalanche are presented in Fig.6. One can see that no loss of x-ray efficiency was observed up to an accumulated charge of.0.1mC/cm$^2$.

**V. Discussions**

**V-1. Aging of gaseous detectors filed with photosensitive vapors**

The processes, which lead to formation of polymer depositions on the detector's electrodes, are very complicated. Basically, they include a dissociation of gas-mixture polyatomic molecules M on fragments F, including ionized ($F^+$) and excited ($F^*$) fragments. Schematically these processes can be illustrated as follows.

1) Dissociation of molecules by an electron (e) impacts and charge and excitation transfer from the atoms A (atoms of noble gases and atoms appearing due to the molecule's dissociation):

$M + e \rightarrow M^+ + e$

$M + e \rightarrow M^* + e$

$M + A* \rightarrow M^{+,*} + A$

$M + A^+ \rightarrow M^{+,*} + A$

$nM + M^{*,+} \rightarrow M_{n+1}^{+,*}$ (polymers)

$M^{+,*} \rightarrow F_1 + F_2$ (and various $F^{+,*}$)

2) Dissociation due to photons (photons of the detecting light and photons from avalanches):

$M + h\nu \rightarrow M^+ + e$

$M + h\nu \rightarrow M^*$

$M^{+,*} \rightarrow F_1 + F_2$ (various $F^{+,*}$).

The main result of these processes is the breakage of chemical bonds and the formation of molecular fragments and radicals (bond braking leads to formation of radicals). These radicals react either with molecules and fragments or with impurities, for example:

F (or $F^{+,*}$) + $M_{impur.}$ $\rightarrow$ polymers.

They may also recombine to each other. Thus there are many various channels for polymer formation. Note, that ionized and excited M ($M^{+,*}$) are much more chemically active than M. They also have larger cross sections of interactions.

Specific reactions for photosensitive vapors TMSi and TMP [23] which leads to polymer formation:

$(CH_3)_4Si + e \rightarrow (CH_3)_4Si^+ \rightarrow (CH_3)_3Si^+ + CH_3$

$TMP^+ + nTMP \rightarrow (TMP)^+_{n+1}$

Even this "very simplified" picture of processes looks rather complicated.

However, in the case of mixtures with photosensitive vapors $Ei_{vapors} < Ei_{main\ mixture}$ and this circumstance allows to simplify the consideration of processes described above. Indeed, through the multistep exchange processes, all excitation and ionization energy will finally be transferred to photosensitive molecules $m_p$ (which may then further dissociate on fragments $m_1$ and $m_2$):

$M^{+,*} + m_p \rightarrow m_p^{+,*} \rightarrow m_1^{+*} + m_2$.

This was experimentally confirmed by light emission measurements, performed for light emission chambers (optical readout mentioned in the introduction) -see fig. 7. As one can see, in mixtures of TEA or TMAE with noble gases or noble gases with mixtures of other molecular gases [24] only emission bands of TEA or TMAE are present in the spectrum. The explanation of this was given in ref. [25]. Thus, in most cases (especially in a low electric field) one should consider only the role of fragments of photosensitive vapors in aging kinematics. This explains why the aging rate depends only slightly on the carrier gas composition, but depends strongly on the nature of the chemical composition of the vapors and the presence of impurities with which the fragments $m_1$ and $m_2$ may react. One can assume that the thickness of the polymer deposition on anode wires depends on the amount of the deposit charge per unit of the anode area. Then for the same total charge, deposit per unit of the anode wire length (usual unites in aging results presentation [14]) the aging rate should be approximately inversely proportional to the anode wire diameter, and this was precisely observed experimentally [14].

**V-2. Aging of solid photocathodes**

**V-2a. CsI photocathode**

As mentioned above, the explanation of the CsI aging mechanism presented in [10] is not satisfactory at all because it is applicable to CsI molecules (vapor phase) only but not to the CsI in solid state phase. In the case of the ideal CsI crystal the absorption of UV photons should lead to the creation of excitons and free electrons in the conduction band (see Fig. 8a.). This would not cause any dissociation.

However, in reality CsI is not an ideal crystal. It contains cracks, vacancies, intersitials, dislocations, chemical impurities, insertions and so on. This creates extra levels schematically shown as "A-levels" and "B-levels" in Fig. 8b. In this picture "A" levels are occupied states above the filled band and B levels are unoccupied states below the conduction band. As a result electron traps are formed and new levels appear, from which the photoemission may occur. Note that CsI photocathode may also may have stohametris deviations (contain excess of Cs or Iodine due to, for example, exposure to light during evaporation or dissociation in water) and this also creates extra-levels in the structure of the energy-bands.

Dramatic and irreversible changes occur when the CsI photocathode is exposed to air. It's initial polycrystalic structure cracks to granules [16]. Absorbed water further damages the photocathode structure and created a "CsI- water" solution in which Cs and

Iodine may dissociate. If an aging test is done with a non-monochromatic light then some contribution to the photoemission may come from the created level and this will enchains the photocurrent. On the other hand, levels can trap free electrons inside the CsI structure and this may cause a charging up effect. Thus there are several competing processes, which contribute to the measured photocurrent. If the aging test is interrupted (for example, the light is blocked) then some relaxation processes may occur [26] and the degraded QE can be partially restored. The same restoration effect one can expect with photocathode heating. Note that both these predictions based on the solid -stated theory [26] were confirmed experimentally [21].

Let's now discuss the possibility of the CsI dissociation by ion impact (see section II-2). Although the positive ions from the avalanche do not have enough kinetic energy to dissociate the CsI (at least at pressure of 1 atm and an electric field <10 kV/cm see for example [27]), they can create insertions and with such a process distract an "ideal" crystalline structure. As a results, the photocathode aging characteristics, especially for photocathodes exposed to air, may depend quite extensively on the photocathode history: cleaness, duration of exposure to air and so on. This may explain a great range of the experimental data mentioned above.

Thus the simplified model described in the section II-2 does not at all reflect the complexity of the aging phenomena. One can therefore speculate that the aging is a result of a gradual destruction of the ideal crystalline structure through the creation of defects, extra levels, insertions, dissociation in water layers with a possible further dissociation on Cs and Iodine atoms. Due to defects electron transfer properties will be diminished and the photocathode, layer by layer, shall be destroyed and as a results the QE will drop.

This model also explains a heating effect- the partial restoration of the QE value due to the so -called relaxation effects [26]. All this allows to bring up the suggestion that the best aging properties should exhibit photocathodes not exposed to air and our experimental data presented in this work exactly supported this assumption.

**V-2b. Aging of photocathodes sensitive to visible light**

Several groups are now able to manufacture gaseous detectors sensitive up to visible light. However, there have been earlier observations from Breskin's group [28] suggesting that such photocathodes have very poor aging properties. Our result show that aging properties of high quality CsI and SbCs photocathode (prepared for example by Hamamatsu) are very good- see Fig 4 and 5. Even when these photocathodes were operated in a gas atmosphere, the ion bombardment did not cause any extra problems. A review of possible mechanisms involved in this photocathode aging can found elsewhere (see for example [22]). In ref. [22] one can also find some discussion concerning the possible dependence between the photocathode resistivity and the variation the QE with time.

**V-2c. Link between the Malter-type effect and breakdowns in gaseous detectors**

For simplicity, in previous considerations of the photocathode aging, we have neglected the role of the deposition on their surfaces due to the gas aging. However, these depositions may play an important role. We will mention here only two major effects at this point:

1) If polymers are formed on the cathode surface, then photoelectrons, created from the cathode, should pass this layer before being able to penetrate to the detector. As a result, some fractions of the electrons will be lost and thus the measured QE will be decreased.
2) The depositions (polymer layers or even absorbed layer of gases) can lead to the Malter-type effect- a sporadic emission of electrons [29]. One should note that a classical theory of the Malter effect suggests a single electron emission [29]. In contrast, however, it was recently discovered that this emission of electrons might occur in the form of a burst or jets of electrons [30]. This effect is very similar to the well-known "explosive" field emission mechanism in vacuum breakdown [31]. Each burst may contain between a few and up to $10^5$ electrons emitted in a time interval from a fraction of μs to ms [29,32]. This Malter-type effect or more precisely, explosive field emission in turn may lead to two other effects:

a) A reduction in the photocathode work function (as well as an increase of electron emission from the photocathode), and therefore an increase in its measured QE;

b) Since these electrons receive full multiplication in gaseous detectors they may cause breakdowns [29,32].

The discussion mentioned above suggests that the gas mixtures in which the photodetectors operate should be carefully chosen: they should not make any deposition ether due to polymerization or any absorbed layer. Note however, that in practice, even in well-selected gas mixtures, it is quite impossible to avoid microscopic dielectric insertions on the cathode surface [31]. This may lead to "explosive" field emission. Thus this type breakdown triggering mechanism looks as a general one [32].

## VI. Conclusions

Photosensitive materials may have a lot applications in various fields, but successes of these applications depend on their aging characteristics. This is why their aging studies are very important.

As a results of the work we can suggest that the one of the most important factors in the photosensitive vapors aging is the system cleanness.

In the case of the solid photocathodes we also found that high quality photocathodes, not exposed to air, have excellent aging properties both in vacuum and the gas atmosphere. This in turn suggests that ion bombardment does not play any important role in the photocathode's aging.

However, polymer depositions and even adsorbed layers when exposed to positive ions bombardment may cause the QE degradation and the Malter –type effect. Thus we can finally conclude that this Malter –type effect (or "explosive" field emission) could be the main breakdown triggering mechanism in any gaseous detectors. This is because in practice it is almost impossible to avoid microscopic dielectric insertions or adsorbed layers on cathode surfaces.

**Figure captions**

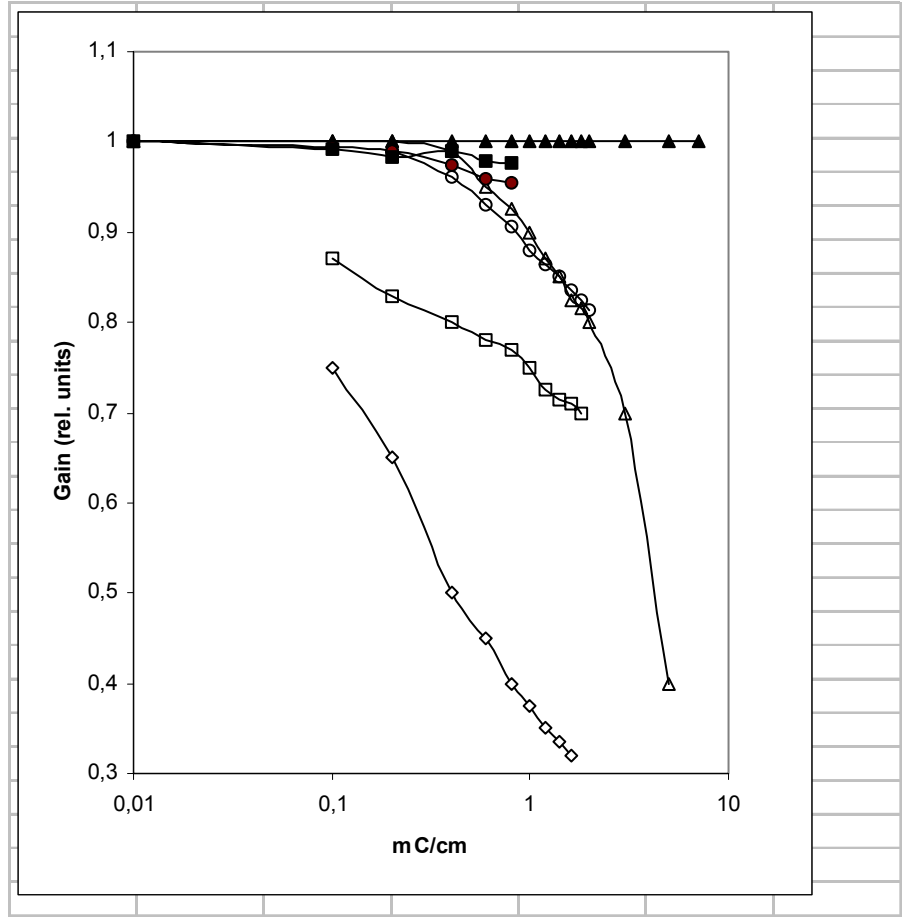

Fig.1 Results of aging tests of several photosensitive vapors: TMAE (open diamonds) in a not very clean gas chamber [14]; TMAE (open squares and triangles) in a clean gas chamber [14]; TEA (open rounds) [14]; EF (filled squares) and TMA (filled rounds)-our data obtained with a sealed SWC (anode diameter of 0.05 mm) at gain of~$10^3$; EF and TMA (filled triangles)-our data obtained with a sealed SWC (anode diameter of 0.2mm), operating in a streamer mode (gain of ~$10^8$).

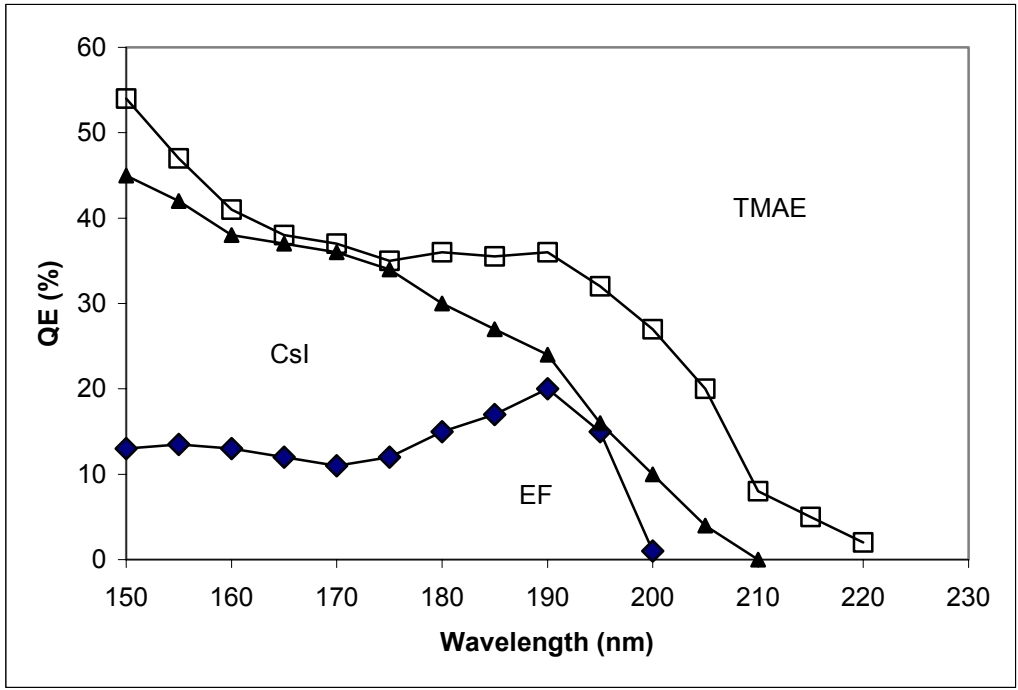

Fig.2 A quantum efficiency of several photosensitive materials

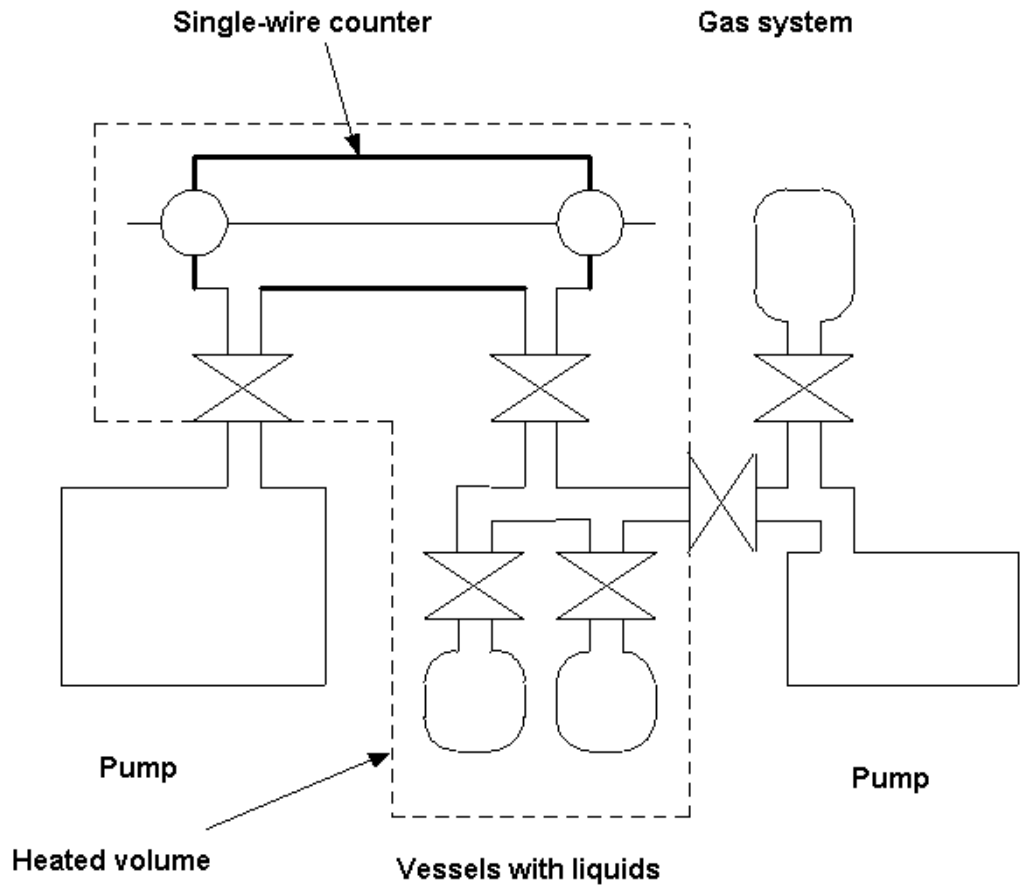

Fig. 3a Experimental set up for manufacturing sealed single-wire counters

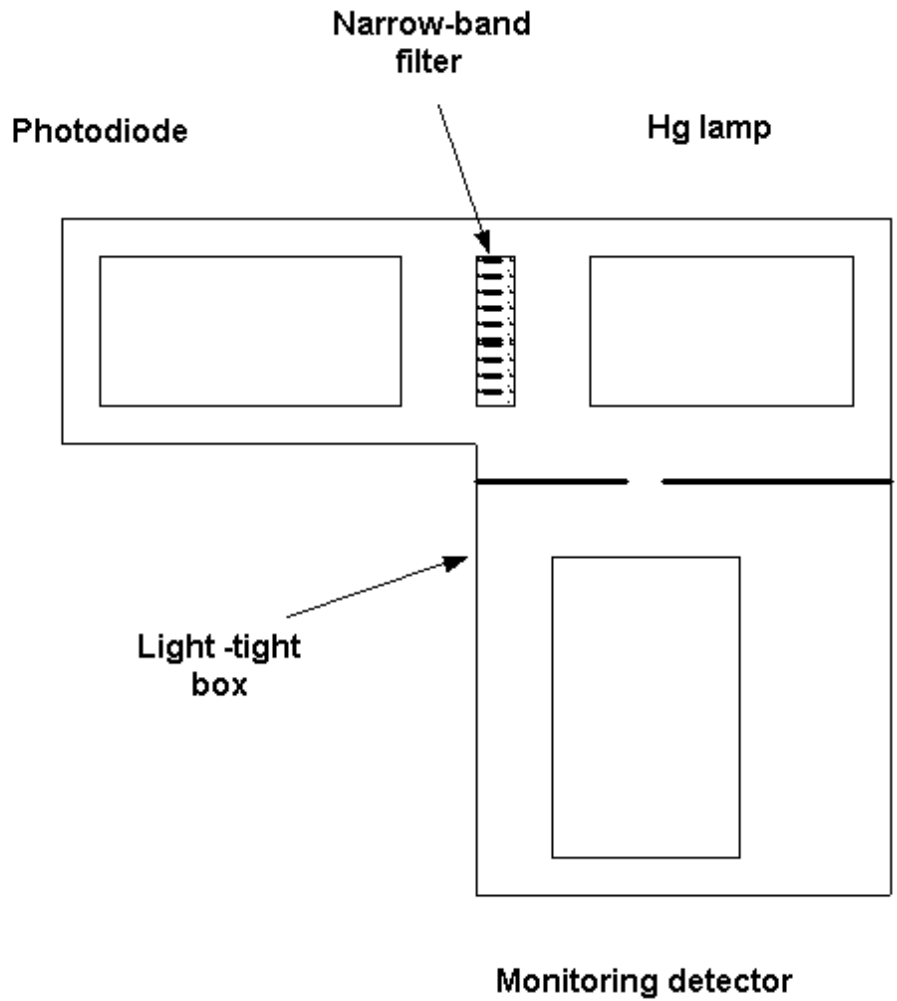

Fig. 3b Experimental set up for photocathodes aging studies

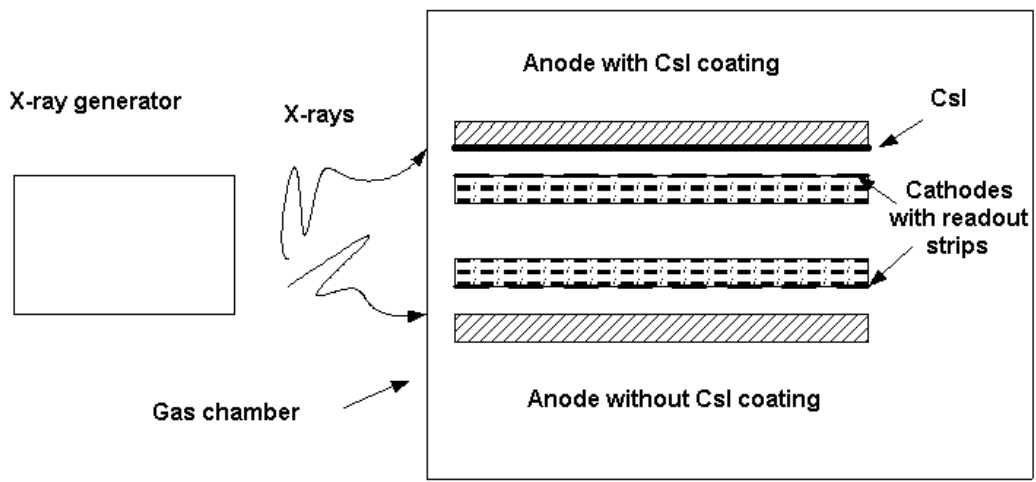

Fig. 3c Experimental set up for a solid X-ray converter aging studies

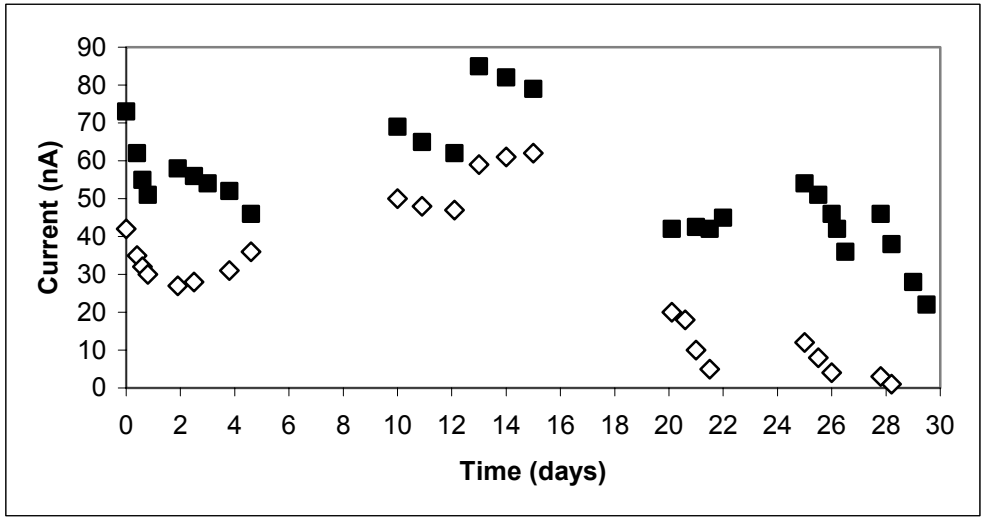

Fig. 4 Results of aging studies of the CsI photocathodes (photocurrent vs. time) manufactured by Hamamatsu: for R1187 (open diamonds) and R1187 filled with Ar (filled squares). Note that there were breaks in measurements between days 5 and 10, 15 and 20, 22 and 25, 26 and 27. One can see that during the breaks the QE value was partially restored.

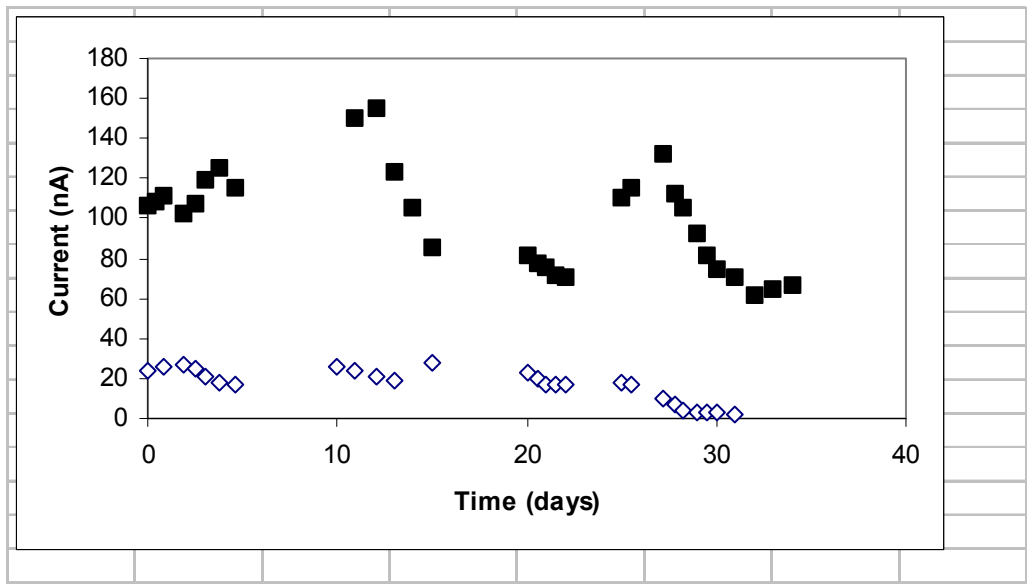

Fig. 5. Results of aging studies of the SbCs photocathodes manufactured by Hamamatsu:R414 (open diamonds) and R250 filled with Ar (filled squares). Note that there were breaks in measurements between days 5 and 10, 15 and 20, 22 and 25, 26 and 27. One can see that during the breaks the QE value was partially restored.

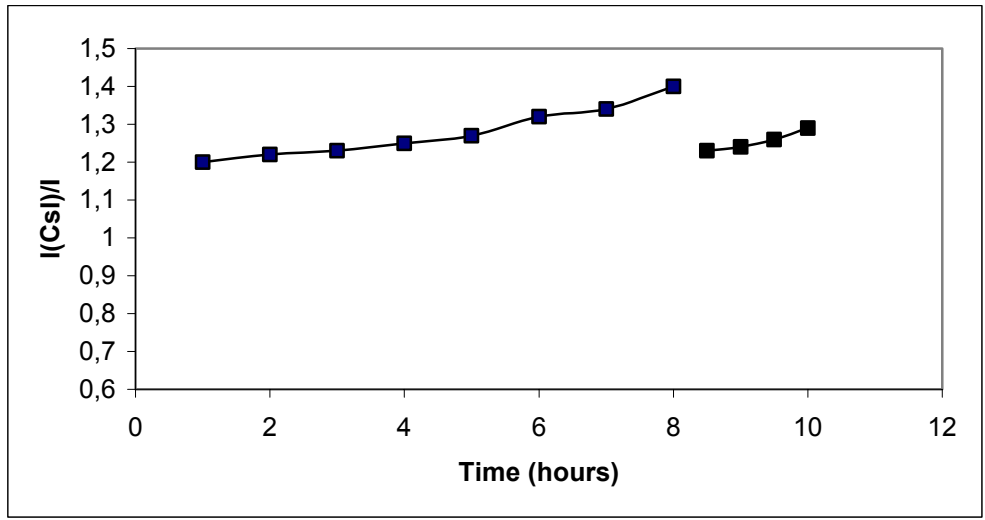

Fig.6 Ratio of two currents (current from the PPAC with the CsI converter divided on the current from the PPAC without any converter) vs. time.. During these measurements the current value from the PPAC without the CsI converter was with good accuracy about 0.1µA. After 8 hours the X-ray flux was blocked for half an hours and then the exposure to the X-rays was continued.

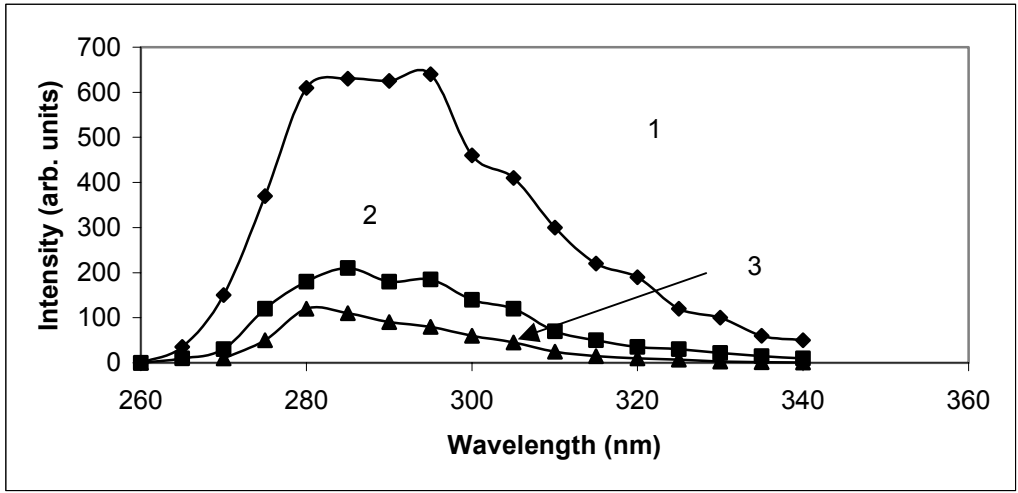

Fig. 7 Emission spectra of various gas mixtures filled with TEA vapors at total pressure of 1 atm:

1) Ar+6%TEA, 2) Kr+6&TEA,3) Xe+6%TEA

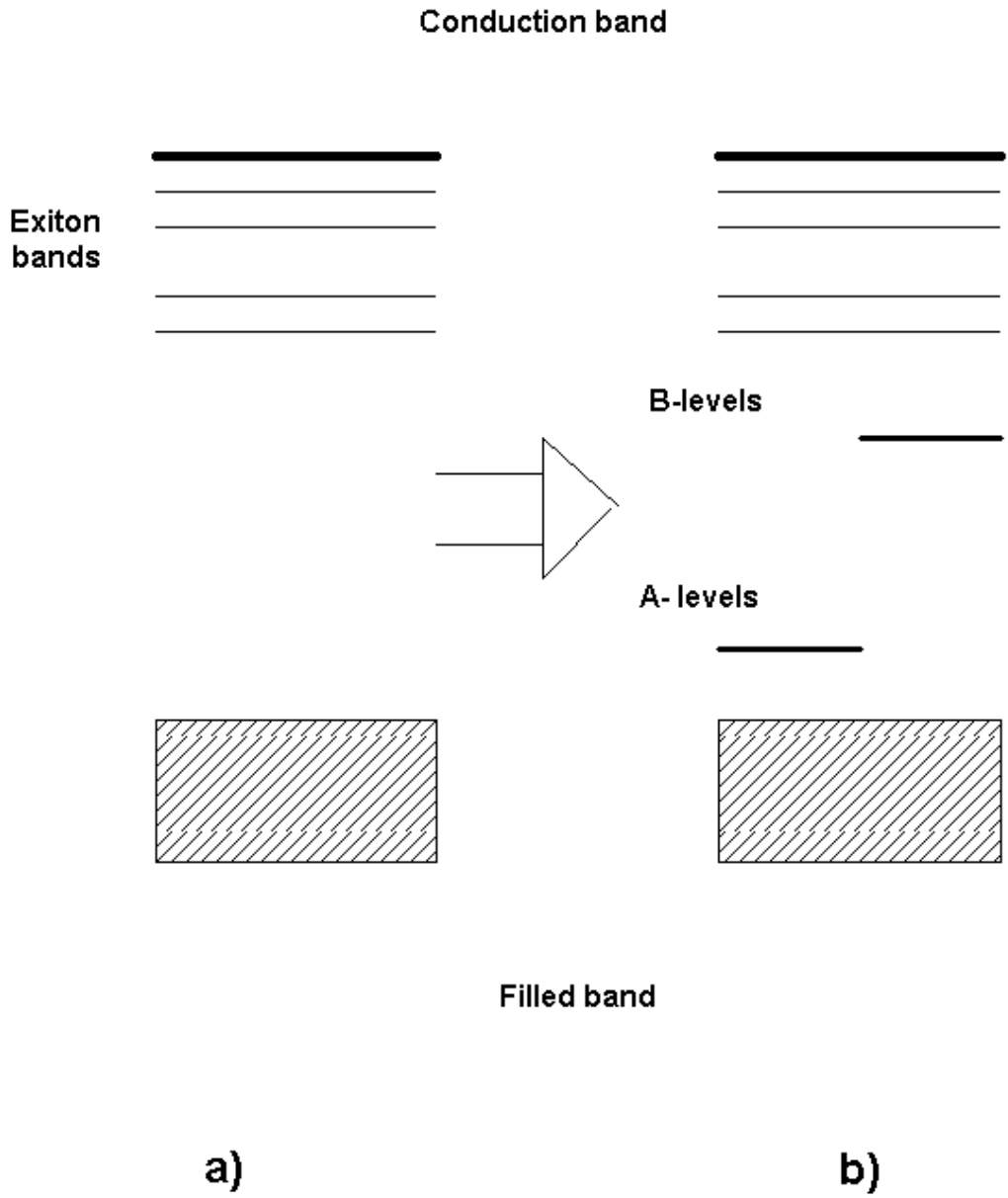

Fig.8 Energy band scheme for an ideal CsI crystal (a) and for the CsI containing various defects (b)